\documentclass[pre,aps,10pt,longbibliography,amsmath,amsfonts,amssymb,twocolumn]{revtex4-1}
\newcommand{\onechi}{0.3}
\newcommand{\twochi}{0.3}
\newcommand{\thrchi}{0.25}
\newcommand{\fouchi}{0.25}
\newcommand{\schi}{0.35}
\newcommand{\echi}{0.4}
\newcommand{\tenchi}{0.4}
\newcommand{\zchi}{0.3}

\newcommand{\oneval}{0.2532(8)}
\newcommand{\twoval}{0.2647(7)}
\newcommand{\thrval}{0.276(1)}
\newcommand{\fouval}{0.2828(8)}
\newcommand{\sval}{0.2855(5)}
\newcommand{\eval}{0.2866(3)}
\newcommand{\tenval}{0.2872(4)}
\newcommand{\zval}{0.2393(3)}

\newcommand{\onepc}{0.572~375(25)}
\newcommand{\twopc}{0.581~85(5)}
\newcommand{\thrpc}{0.5905(1)}
\newcommand{\foupc}{0.5983(1)}
\newcommand{\spc}{0.6112(1)}
\newcommand{\epc}{0.621~11(1)}
\newcommand{\tenpc}{0.628~75(5)}
\newcommand{\zpc}{0.562~142(3)}
\usepackage{graphicx}
\usepackage{CJK}
\begin{document}
\begin{CJK*}{UTF8}{mj}
\title{Branching annihilating random walks with long-range attraction in one dimension}
\author{Su-Chan Park (박수찬)}
\affiliation{Department of Physics, The Catholic University of Korea, Bucheon 14662, Republic of Korea}
\begin{abstract}
We introduce and numerically study the branching annihilating random walks with long-range attraction (BAWL).
The long-range attraction makes hopping biased in such a manner that
particle's hopping along the direction to the nearest particle has larger
transition rate than hopping against the direction.
Still, unlike the L\'evy flight, a particle only hops to one of its nearest-neighbor sites.
The strength of bias takes the form $x^{-\sigma}$ with non-negative $\sigma$, where
$x$ is the distance to the nearest particle from a particle to hop.
By extensive Monte Carlo simulations, we show that the critical decay exponent $\delta$ 
varies continuously with $\sigma$ up to $\sigma=1$ and $\delta$ is the same
as the critical decay exponent of the directed Ising (DI) universality class for $\sigma \ge 1$.
Investigating the behavior of the density in the absorbing phase, we argue that
$\sigma=1$ is indeed the threshold that separates the DI and non-DI critical behavior.
We also show by Monte Carlo simulations that branching bias with symmetric hopping 
exhibits the same critical behavior as the BAWL.
\end{abstract}
\date{\today}
\maketitle
\end{CJK*}
\section{Introduction}
The branching annihilating random-walks model (BAW)~\cite{TT1992} is a reaction-diffusion system 
with pair annihilation [$2A\rightarrow 0$] and branching $m$ offspring by a particle [$A\rightarrow (m+1)A$]
as well as (symmetric) diffusion. The competition between pair annihilation and branching 
can bring about an absorbing phase transition between an active phase with nonzero steady-state density
and an absorbing phase with zero steady-state density. 
The BAW exhibits rich phenomena in that critical behavior depends on
the parity of the number $m$ of offspring~\cite{TT1992,J1994,Zhong1995,KP1995}.
It belongs to the directed percolation (DP) 
universality class~\cite{BH1957,GT1979,CS1980,J1981,G1982} for odd $m$, 
whereas it belongs to the directed Ising (DI) universality 
class~\cite{GKvdT1984,KP1994,MO1996,CT1996,H1997,Canet2005,HCDM2005} for even $m$.
For a review of these two classes, see, e.g., Refs.~\cite{H2000,O2004,BookHHL2008}

When a global hopping bias is introduced to the BAW in such a way that
hopping along a predefined direction is preferred (for example,
in one dimension hopping to the right has larger transition rate
than hopping to the left), 
this bias in the (asymptotic) field theory is gauged away by a Galilean 
transformation~\cite{PHP2005a} and, in turn, critical behavior is not affected by the global bias.
Recently, a local hopping bias was introduced to the BAW~\cite{DR2019} in such a manner that 
a particle prefers hopping toward the nearest particle. 
Since a particle is likely to get close to the nearest particle by the local bias,
this form of interaction associated with the local bias is termed as attraction in Ref.~\cite{Park2020A}.
Since hopping along any direction is equally likely on average,
no macroscopic current is produced by the local bias.
In this sense, the Galilean transformation cannot remove the local bias
and, in turn, the local bias can be relevant in the renormalization-group (RG) sense.
Indeed, it was shown that the local bias changes the critical behavior when the number $m$ of offspring
is even~\cite{DR2019,Park2020A}.

Unlike a long-range jump (L\'evy flight) introduced to models exhibiting an absorbing
phase transition~\cite{Mollison1977,Book:G1986}, 
every particle still hops to one of its nearest-neighbor sites.
In this sense, one may think of the local bias as short-range interaction. This idea seems to have support 
because the BAW with an odd number of offspring is not affected by
the local bias, while L\'evy flight applied
to DP models changes critical behavior~\cite{Janssen1999,HH1999,JS2008}.
However, it was argued that the local bias is irrelevant (in the RG sense) in the DP class not because
the bias is short-ranged but because
spontaneous annihilation ($A\rightarrow 0$) arising by combination
of branching with pair annihilation ($A\rightarrow 2A \rightarrow 0$) 
removes the long-range nature of the local bias for odd $m$~\cite{Park2020A}.

To reveal clearly the long-range nature of the local bias for the case of even number of offspring,
Ref.~\cite{Park2020A} studied a modified model by introducing the range $R$ of attraction.
In the modified model, a particle is attracted to the nearest particle only if
the distance between the two particles is not larger than $R$.
When $R$ is finite, the model with even $m$ turned out to crossover to the DI class  
and the crossover behavior for large $R$ is described by the exponent $\phi$, 
which is found to be $1.39\pm 0.04$~\cite{Park2020A}.
Therefore, it is concluded that the different critical behavior from the DI class in Ref.~\cite{DR2019} 
is attributed to the long-range nature of the local bias. 

Since long-range interaction usually entails continuously varying critical 
exponents~\cite{Janssen1999,HH1999,JS2008}, it is natural to ask 
if the local bias with appropriate generalization can trigger continuously varying exponents.
The aim of this paper is to answer this question by studying such a generalized model that
the strength of the local bias depends on the distance $x$ to the nearest particle by a power-law 
function $x^{-\sigma}$. The case with $\sigma=0$ will correspond to the model in Ref.~\cite{DR2019}.
We will investigate how the critical behavior changes with the value of $\sigma$.

The structure of this paper is as follows. In Sec.~\ref{Sec:model}, we define a model with a local bias.
As explained above, the strength of the bias becomes a power-law function of 
distance to the nearest particle.
We will call this model the branching annihilating random walks with long-range attraction 
(BAWL).
In Sec.~\ref{Sec:res}, we present our simulation results,
focusing on the critical decay exponent that is defined in Sec.~\ref{Sec:model}.  
We will also find $\sigma_c$ that separates 
the DI critical behavior (for $\sigma \ge \sigma_c$) and non-DI critical behavior (for $\sigma < \sigma_c$).
In Sec.~\ref{Sec:dis}, we discuss what happens if branching is biased.
Section~\ref{Sec:sum} summarizes the paper.

\section{\label{Sec:model}Model and Methods}
The BAWL is defined on a one-dimensional lattice of size $L$ with
periodic boundary conditions. 
Each site $i$ ($i=1,2,\ldots,L$) is characterized by an occupation number $a_i$ that takes either one
or zero. If $a_i=1$, we say that there is a particle at site $i$.
If $a_i=0$, we say that site $i$ is vacant.
For later purpose, we define $r_i$ and $l_i$ such that
\begin{align}
	\nonumber
	r_i &= \min\left\{ x | a_{i+x}=1,\; x>0\right \},\\
	l_i &= \min\left\{ x | a_{i-x}=1,\;x>0\right \},
\end{align}
where we assume that site $j+L$ is identical to site $j$ (periodic boundary condition).
In words, $r_i$ ($l_i$) is the distance from site $i$ to the nearest particle on the right-hand (left-hand) side.
%One can regard $a_i$, $r_i$, $l_i$ as random variables.

If there is a particle at site $i$ ($a_i=1$), it either hops to one of its nearest-neighbor sites
with rate $p$ (hopping event) or
branches four offspring with rate $1-p$ (branching event).
In the hopping event, it hops to site $i\pm 1$  with probability $q_\pm$, where
\begin{align}
	q_\pm = \frac{1}{2} \pm \zeta x^{-\sigma},\quad x = \min\{r_i,l_i\},\quad \sigma \ge 0,
	\label{Eq:q}
\end{align}
with ($0\le \epsilon \le 0.5$)
\begin{align}
	\label{Eq:zeta}
	\zeta 
	&= \begin{cases}
		\epsilon, &\text{ if } r_i < l_i,\\
		-\epsilon, &\text{ if } r_i > l_i,\\
		0, &\text{ if } r_i = l_i.
	\end{cases}
\end{align}
Notice that $q_\pm$ mimics attraction by the nearest particle.

In the branching event, its four offspring are placed at sites
$i-2$, $i-1$, $i+1$, and $i+2$ ($A\rightarrow 5A$).
If a particle is to be placed at an already occupied site either by hopping or branching,
these two particles are annihilated immediately ($2A\rightarrow 0$).
We summarize the above dynamic rules as follows:
\begin{subequations}
	\label{Eq:rules}
\begin{eqnarray}
	1_i a_{i+ 1} \rightarrow 0_i\, \overline{a}_{i+ 1}
	&\text{ rate }& p q_+,\\
	a_{i- i} 1_i \rightarrow  \overline{a}_{i- 1}\,0_i
	&\text{ rate }& p q_-,\\
	1_i\,a_{i\pm 1}\,a_{i\pm2}  \rightarrow 1_i\, \overline{a}_{i\pm 1}\,\overline{a}_{i\pm 2} &\text{ rate }& 1-p,
\end{eqnarray}
\end{subequations}
where $1_i$ ($0_i$) means that $a_i$ is one (zero) 
and $\overline{a}_j \equiv 1 - a_j$.
We set $\epsilon=0.1$ in simulations but other choice of nonzero $\epsilon$ does not change our conclusion.

The algorithm we have used to simulate the corresponding master equation to the rule~\eqref{Eq:rules} is as follows.
Assume that there are $N_t$ particles at time $t$.
We choose one particle among $N_t$ particles at random with equal probability.
The chosen particle branches four offspring with probability $1-p$ or
hops toward (against) the nearest particle with probability $pq_+$ ($pq_-$), 
where $q_\pm$ is defined in Eq.~\eqref{Eq:q}.
If two particles happen to occupy a site, these two particles are removed in no time.
After the change, time increases by $1/N_t$.

The BAWL with $\sigma=0$, which is identical to the model in Ref~\cite{DR2019}, does not
belong to the DI class, while 
the BAWL under $\sigma \rightarrow \infty$ limit is equivalent to the model in Ref.~\cite{Park2020A}
with $R=1$ and, in turn, belongs to the DI class.
Thus, there should be $\sigma_c$ such that the BAWL with $\sigma\ge \sigma_c$ belongs
to the DI class.  In this paper, we will 
find $\sigma_c$ and investigate the critical behavior for $\sigma<\sigma_c$.

We will study the average density $\rho$ of occupied sites
at time $t$ defined as
\begin{align}
	\rho(t) = \frac{1}{L} \sum_{i=1}^L \langle a_i \rangle,
\end{align}
where $\langle \cdots \rangle$ stands for average over all ensemble.
The configuration with $a_i=1$ for all $i$ will be used as an initial condition in this paper.

At the critical point, $\rho(t)$ is expected to show a power-law behavior with a critical decay exponent
$\delta$ such that
\begin{align}
	\rho(t) = A t^{-\delta} \left [ 1 + Bt^{-\chi} + o(t^{-\chi}) \right ],
\end{align}
where $t^{-\chi}$ is the leading term of corrections to scaling, 
$o(x)$ stands for all terms that decrease faster than
$x$ as $x\rightarrow 0$, and $A$, $B$ are constants.
We will call $\chi$ the corrections-to-scaling exponent.

To find $\delta$, we study an effective exponent $-\delta_\text{e}$ defined as
\begin{align}
	-\delta_\text{e}(t,b) \equiv \frac{\ln [ \rho(t)/\rho(t/b) ]}{\ln b},
\end{align}
where $b$ is a constant.
At the critical point, the effective exponent in the long time limit should behave as
\begin{align}
	-\delta_\text{e}(t,b) \approx -\delta - B \frac{b^\chi-1}{\ln b} t^{-\chi}.
	\label{Eq:delasym}
\end{align}

From Eq.~\eqref{Eq:delasym}, it is obvious that at the critical point
$-\delta_\text{e}$, when treated as a function of $t^{-\chi}$, should show a linear behavior for small $t^{-\chi}$.
On the other hand, if the system is slightly off the 
critical point and is actually in the active (absorbing) phase,
$-\delta_\text{e}$ should eventually veer up (down) as $t^{-\chi} \rightarrow 0$.
Accordingly, we can find the critical point by observing how $-\delta_\text{e}$ behaves.
Once we find the critical point, the critical decay exponent can be found by linear
extrapolation of $-\delta_\text{e}$ vs $t^{-\chi}$ at the critical point.

To estimate $\delta$ accurately, information of $\chi$ is crucial.
To find $\chi$, we analyze a corrections-to-scaling function $Q$
defined as~\cite{Park2013,Park2014PRE}
\begin{align}
	Q(t;b,\chi) = \frac{\ln \rho(t/b^2) + \ln \rho(t) - 2 \ln \rho(t/b)}{(b^\chi-1)^2},
\end{align}
whose asymptotic behavior at the critical point is $Q\sim B t^{-\chi}$ regardless of the value of $b$
if $\chi$ is correctly chosen. Notice that if $B$ is positive (negative), $-\delta_\text{e}$ approaches $-\delta$ 
from below (above). In our system, we actually found that $B$ is negative.

For convenience, an $i$th measurement is performed at time $T_i$ defined as
\begin{align}
	T_i = 
	\begin{cases} 
		i , & i\le 40,\\
		\lfloor 40 \times 2^{(i-40)/15} \rfloor, & 41 \le i \le 55,\\
		2 T_{i-15}, & 56 \le i,
	\end{cases}
\end{align}
where $\lfloor x \rfloor$ is the floor function (greatest integer not larger than $x$).
With this choice of measurement time, 
we can set $b=2^{n}$ ($n=1,2,\ldots$) to analyze
the effective exponent as well as the corrections-to-scaling function.

\section{\label{Sec:res}Results}
In this section, we present our simulation results for
the critical decay exponent $\delta$ for various values of $\sigma$.
To begin, we analyze the BAWL with $\sigma=0.1$ and 0.3.
In simulations for these two cases, the system size is $L=2^{23}$ and the maximum observation time is
$T_{289} \approx 4 \times 10^6$.
The number of independent runs is between 80 and 200.
We first analyzed the corrections-to-scaling function $Q$ and we found $\chi$ to be 
0.3 and 0.25 for $\sigma=0.1$ and 0.3, respectively,see Supplemental Material~\cite{supp}.
In Fig.~\ref{Fig:e2}, we depict the effective exponent as a function
of $t^{-\chi}$ for $\sigma=0.1$ [Fig.~\ref{Fig:e2}(a)] and 0.3 [Fig.~\ref{Fig:e2}(b)] with 
$b=16$.
\begin{figure}
	\includegraphics[width=\linewidth]{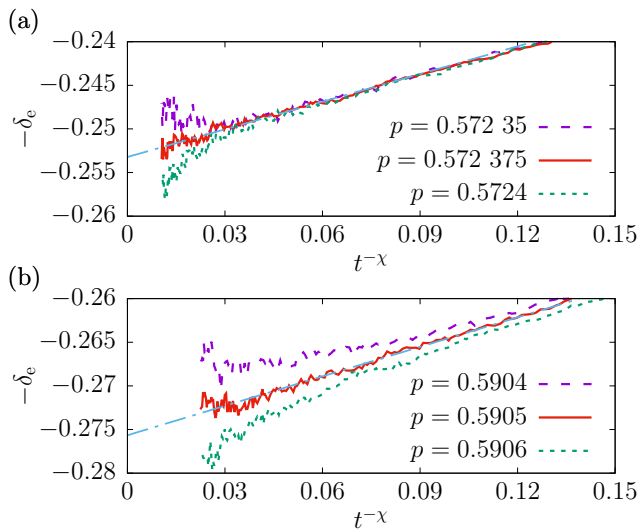}
	\caption{\label{Fig:e2} Plots of $-\delta_\text{e}$ vs $t^{-\chi}$
	(a) for $\sigma=0.1$ at $p=0.573~35$, $0.572~375$, $0.5724$ (top to bottom) with $\chi=0.3$ and $b=16$
	and (b) for $\sigma=0.3$  at $p=0.5904$, 0.5905, 0.5906 (top to bottom) with $\chi=0.25$ and $b=16$. 
	The (dot-dashed cyan) straight lines overlapping with the middle curves show the results of 
	linear extrapolation for the critical decay exponent.
	Clearly, the critical decay exponent $\delta$ varies with $\sigma$.
	}
\end{figure}

Since middle curves in both panels show linear behaviors,
while the other curves eventually veer up or down,
we estimate the critical point as $p_c = 0.572~375(25)$ for $\sigma=0.1$
and $p_c = 0.5905(1)$ for $\sigma=0.3$, where
the numbers in parentheses indicate uncertainty of the last digits.
By linear extrapolation,
we get $\delta = \oneval$ for $\sigma=0.1$ and $\thrval$ for $\sigma=0.3$.
It is clear that $\delta$ does depend on $\sigma$, which
is a typical feature of absorbing phase transitions with long-range jump~\cite{Janssen1999,HH1999,Vernon2001,JS2008}. Once again we confirm the claim in Ref.~\cite{Park2020A} that
the model with hopping bias in Ref.~\cite{DR2019} does not belong to the DI class because 
of long-range interaction.
\begin{figure}
	\includegraphics[width=\linewidth]{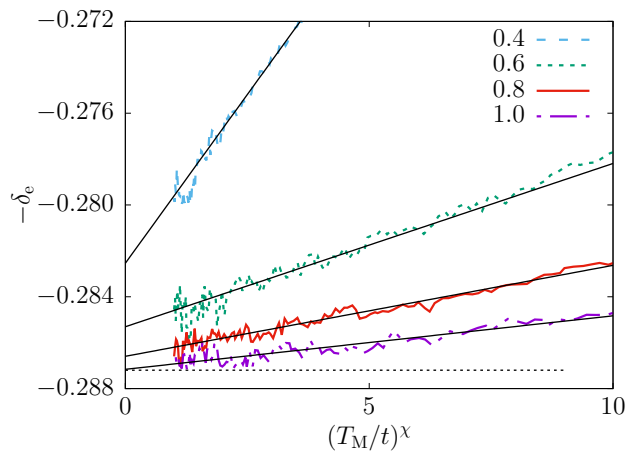}
	\caption{\label{Fig:da} Plots of $-\delta_\text{e}$ vs $(T_M/t)^\chi$
	with $b=32$ at the critical point 
	for $\sigma=0.4$, 0.6, 0.8 and 1 (top to bottom),
	where $T_M$ is the maximum observation time. 
	Here, $T_M =T_{289}\approx 4\times 10^6$ for $\sigma=0.4$ and
	$T_M =T_{309}\approx 10^7$ for other cases.
	Straight lines are results of linear extrapolation and
	the dotted horizontal line indicates the critical decay exponent of the
	DI class.
	}
\end{figure}

\begin{table}[b]
	\caption{\label{Table:expo} Critical points ($p_c$), corrections-to-scaling exponents ($\chi$), 
	and critical decay exponents ($\delta$) of the BAWL. The
	numbers in parentheses indicate uncertainty of the last digits.}
	\begin{ruledtabular}
		\begin{tabular}{llll}
			$\sigma$&$p_c$&$\chi$&$\delta$\\
			\hline
			0\footnotemark[1]  &\zpc  &\zchi&\zval  \\
			0.1&\onepc&\onechi&\oneval\\
			0.2&\twopc&\twochi&\twoval\\
			0.3&\thrpc&\thrchi&\thrval\\
			0.4&\foupc&\fouchi&\fouval\\
			0.6&\spc &\schi &\sval  \\
			0.8&\epc  &\echi&\eval \\
			1.0&\tenpc &\tenchi &\tenval 
		\end{tabular}
	\end{ruledtabular}
	\footnotetext[1]{From Ref.~\cite{Park2020A}.}
\end{table}

We have established that the critical decay exponent varies with $\sigma$. Now,
we move on to finding $\sigma_c$.
Recall that the BAWL with $\sigma\ge \sigma_c$ is supposed to belong to the DI class.
We simulated the system of size $L=2^{23}$ for various $\sigma$'s.
As we have done in Fig.~\ref{Fig:e2}, we first found
$\chi$ and $p_c$, then analyzed the effective exponent, see Supplemental Material~\cite{supp}.

Figure~\ref{Fig:da} depicts the resulting effective exponents at the critical point for
$\sigma = 0.4$, 0.6, 0.8, and 1 against $(T_M/t)^\chi$,
where $T_M$ is the maximum observation time of each simulation for the 
corresponding parameter set.
When $\sigma < 0.8$, the estimate of $\delta$ is clearly distinct from $\delta$ of the DI class
that is shown as a dotted horizontal line in Fig.~\ref{Fig:da}.
For $\sigma=1$, the critical decay exponent is hardly discernible from $\delta$ of the DI class, which
seems to suggest $\sigma_c=1$. Our preliminary simulations also showed that $\delta$ remains the same
for $\sigma > 1$ (not shown here).

To affirm that $\delta$ for the case of $\sigma=0.8$ is indeed larger than the critical decay exponent of the DI 
class, we extensively performed simulations for this case (800 independent runs are averaged).
As shown in Fig.~\ref{Fig:da},
our simulation results suggest that $\sigma_c$ is indeed larger than 0.8, see Supplemental Material~\cite{supp}.

\begin{figure}
	\includegraphics[width=\linewidth]{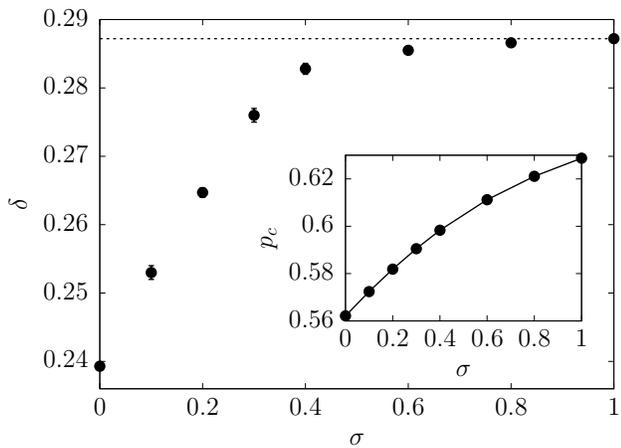}
	\caption{\label{Fig:fin} Plot of $\delta$ vs $\sigma$.
	The critical decay exponent of the DI class is shown as a horizontal
	dotted line.
	The size of the error bar is comparable to the symbol size. 
	(Inset) Plot of $p_c$ vs $\sigma$. The line is for guides to the eyes.
	}
\end{figure}

The values of $p_c$, $\chi$, and $\delta$ for various $\sigma$'s~\cite{supp}
are summarized in Table~\ref{Table:expo} and
in Fig.~\ref{Fig:fin}, we graphically show how $\delta$ and $p_c$ depend on $\sigma$.

Now we will argue that $\sigma_c$ is indeed one.
Since the DI class is intimately related to the annihilation fixed point~\cite{CT1996,Canet2005},
a necessary condition for a model to belong to the DI class is that 
the asymptotic behavior of density  
should be $t^{-0.5}$ in the absorbing phase.
In this context, we will analyze how the density of the BAWL with $p=1$ 
(without branching) behaves in the long time limit.

In the absorbing phase, the density approaches zero as $t\rightarrow \infty$. Hence,
the asymptotic behavior of the density for the BAWL with $p=1$ 
can be understood by studying a random walk model with an attracting center at the origin.
In this random walk model,
a walker located at site $n$ ($n>0$) hops to the right with
rate $(1 - v n^{-\sigma})/2$ and to the left with rate $(1+v n^{-\sigma})/2$.
Now we will find the mean first-passage time to the origin, 
once it starts from site $m$.
It is convenient to regard the origin as an absorbing wall.

The analysis starts from writing down the master equation ($n\ge 1$)
\begin{align}
	\frac{\partial}{\partial t} P_n(t)
	=&- P_n(t)+ \frac{1 + v (n+1)^{-\sigma}}{2} P_{n+1}(t) \nonumber \\
	&+ \frac{1 - v (n-1)^{-\sigma}}{2} (1-\delta_{n,1})P_{n-1}(t),\label{Eq:master1} \\
	\frac{\partial}{\partial_t} P_0(t) = &\frac{1+v}{2} P_1(t),
\end{align}
where $P_n(t)$ is the probability that the walker is at site $n$ at time $t$.
For $n\ge 2$, we rewrite Eq.~\eqref{Eq:master1} as
\begin{align}
	\frac{\partial}{\partial t}P_n(t)
	= - \partial_n \left [ -vn^{-\sigma} P_n(t)
	\right ] +\frac{1}{2} \partial_n^2 P_n(t) ,
\end{align}
where $\partial_n f(n) \equiv [f(n+1)-f(n-1)]/2$ and 
$\partial_n^2 f(n) \equiv f(n+1) + f(n-1) - 2 f(n)$.
Taking (naive) continuum limit, we get a Fokker-Planck equation ($n$ is now a continuous variable)
\begin{align}
	\frac{\partial}{\partial t}P(n,t) = - \frac{\partial}{\partial n} \left [ -v n^{-\sigma} P(n,t)\right ]
	+ \frac{1}{2} \frac{\partial^2}{\partial n^2} P(n,t),
\end{align}
which is equivalent to the Langevin equation
\begin{align}
	\dot n = -v n^{-\sigma} + \xi,
\end{align}
where $\xi$ is the white noise with zero mean and unit variance.
\begin{figure}
	\includegraphics[width=\linewidth]{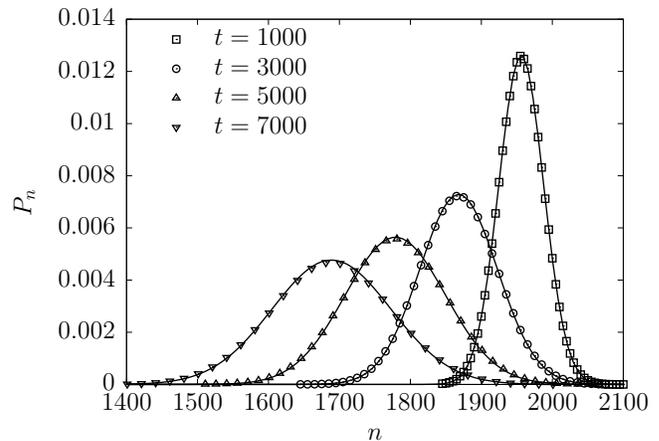}
	\caption{\label{Fig:pr} Plots of $P_n(t)$ vs $n$ at
	$t=1000$, 3000, 5000, and 7000 (right to left).
	Initial position is set $m=2000$.
	Solid curves depicts the approximate solution~\eqref{Eq:Prapp}.
	}
\end{figure}

Using a mean-field-like approximation $\langle n^{-\sigma} \rangle \approx \langle n \rangle^{-\sigma}$,
where $\langle \cdots \rangle$ is the average over noise,
we get
\begin{align}
	\langle \dot n \rangle \approx - \frac{v}{ \langle n \rangle^{\sigma}}\Rightarrow
	\langle n \rangle \approx m \left [ 1 - \frac{(1+\sigma)vt}{m^{1+\sigma}} \right ]^{1/(1+\sigma)},
\end{align}
where $m$ is the initial position of the walker. 
If we further assume that the white noise makes $P_n$ be a Gaussian with variance $t$, we arrive at
\begin{align}
	P_n(t) \approx \frac{1}{\sqrt{2\pi  t}} \exp
	\left [ - \frac{(n - \langle n \rangle)^2}{2  t} \right ],
	\label{Eq:Prapp}
\end{align}
for sufficiently large $n$ (and $m$). 

To check how good the approximation is,
we performed Monte Carlo simulations for the
continuous time master equation~\eqref{Eq:master1} with $\sigma=0.2$, $v=0.2$, and $m=2000$.
In Fig.~\ref{Fig:pr}, we show $P_n(t)$ at $t=1000$, 3000, 5000, 7000
together with Eq.~\eqref{Eq:Prapp}. 
Our approximation is in an excellent agreement with numerical (exact) result.

If $\sigma<1$, the mean first-passage time $\tau$ to the origin is obtained by 
$\langle n \rangle=0$, which gives $\tau \sim m^{1+\sigma}$. 
On the other hand, if $\sigma>1$, the spreading by fluctuation is faster than 
the deterministic motion. Accordingly,
time $\tau$ to arrive at the origin is dominated by diffusion, which gives $\tau \sim m^{2}$.
If we write $\tau \sim m^z$, we find
\begin{align}
	z = \begin{cases}
		1+\sigma, &\sigma<1\\
		2, &\sigma \ge 1.
	\end{cases}
	\label{Eq:taum}
\end{align}

\begin{figure}
	\includegraphics[width=\linewidth]{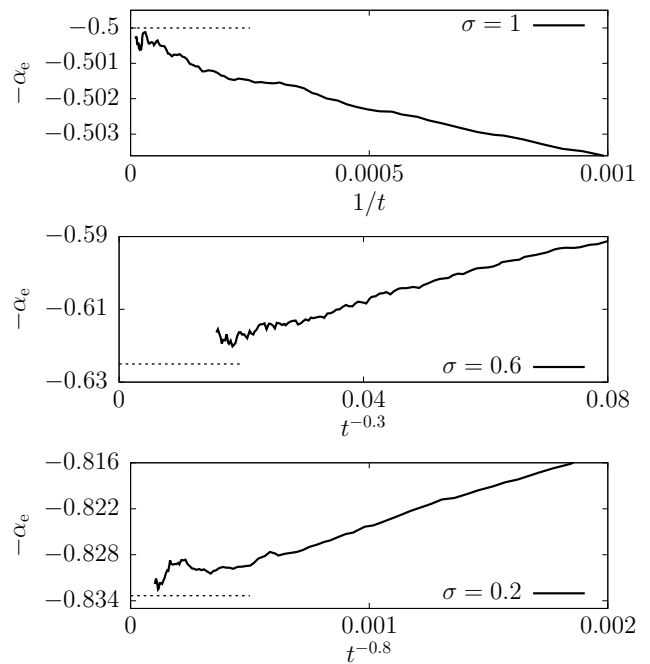}
	\caption{\label{Fig:AA} Plots of $-\alpha_\text{e}$ vs $t^{-\chi}$ for
	$\sigma=0.2$ ($\chi=0.8$: bottom), $\sigma=0.6$ ($\chi=0.3$: middle), and
	$\sigma=1$ ($\chi=1$: top).
	Dotted line segments indicate the anticipated value of $-\alpha$ from Eq.~\eqref{Eq:al}
	}
\end{figure}

From Eq.~\eqref{Eq:taum} and the scaling argument for the pair annihilation dynamics~\cite{KR1984,KR1985},
we predict that the long time behavior of the density is $t^{-\alpha}$ with
\begin{align}
	\alpha = 1/z =
	\begin{cases}
		1/(1+\sigma), & \text{ if } \sigma < 1,\\
		1/2, & \text{ if } \sigma \ge 1.
	\end{cases}
	\label{Eq:al}
\end{align}
To confirm the anticipation, we simulated the BAWL with $\epsilon=0.1$ and $p=1$ for 
various $\sigma$'s.
We present the behavior of effective exponent $-\alpha_\text{e}$ for $\sigma=0.2$,
0.6, and 1 in Fig.~\ref{Fig:AA}, which shows an excellent agreement with the analytic 
argument.

From the above analysis, the BAWL with $\sigma<1$ should not belong 
to the DI class, as we have seen in Fig.~\ref{Fig:da}. 
Since the BAWL with $\sigma=1$ belongs to the DI class as shown in Fig.~\ref{Fig:da},
we conclude that the upper bound $\sigma_c$ is indeed 1

\section{\label{Sec:dis} Discussion: Branching bias}
We have shown that the local hopping bias due to long-range attraction with decreasing strength
as $x^{-\sigma}$ continuously changes the critical decay exponent of the BAWL when $\sigma \le 1$.
Now, we would like to ask which one determines the critical behavior,
\emph{hopping} bias or \emph{bias in itself}. To answer this question, we modify the BAWL 
in such a way that hopping is symmetric but branching is biased.
To be concrete, we will now investigate a model with dynamics
\begin{subequations}
	\label{Eq:modi}
\begin{eqnarray}
	1_i a_{i\pm 1} \rightarrow 0_i\, \overline{a}_{i+ 1}
	&\quad\text{ rate }& p/2,\\
	1_i\prod_{k=1}^4 a_{i+ k} \rightarrow 1_i\,\prod_{k=1}^4 \overline{a}_{i+ k} &\quad\text{ rate }& (1-p) q_+,\\
	1_i\prod_{k=1}^4 a_{i- k} \rightarrow 1_i\,\prod_{k=1}^4 \overline{a}_{i- k} &\quad\text{ rate }& (1-p) q_-,
\end{eqnarray}
\end{subequations}
where $q_\pm$ is the same as in Eq.~\eqref{Eq:q} and we use the same notation as in Eq.~\eqref{Eq:rules}.

Before presenting simulation results, let us ponder on what would happen in this modified model.
The driven pair contact process with diffusion (DPCPD)~\cite{PHP2005a} would be a good starting point 
for our discussion. In the DPCPD, though it has global bias, 
only presence of bias is an important factor to determine the universality class, as 
it is immaterial whether hopping or branching is biased~\cite{PP2009}. 
In this regard, one would conclude that \emph{bias in itself} is relevant (in the RG sense) and that 
the critical behavior of the BAWL would not be affected by to which dynamic process the local bias is applied.
However, the DPCPD should be considered a system with two independent fields
and both the hopping bias and the branching bias in the DPCPD generates a relative bias 
between the two fields~\cite{PHP2005a,PP2008EPJB}.
Since the BAW is described by a single field~\cite{CT1996,Canet2005}, the discussion about the DPCPD
would not give a clear answer to our question.

In the mean time, one may easily come up with an argument that only \emph{hopping} bias is relevant,
because the density of the modified model with $p=1$ (trivially) behaves as $t^{-0.5}$ for any $\sigma$.
This should be compared with the discussion in Sec.~\ref{Sec:res}, based on the analysis of the BAWL with $p=1$.
However, this argument has a serious flaw; 
the dynamics at $p=1$ may not represent the absorbing phase of the modified model.
An example in this context is the BAW with one offspring (BAW$_1$). 
As in the BAWL, let us denote the branching rate of the BAW$_1$ by $1-p$.
If $p=1$, the density (again trivially) decays as $t^{-0.5}$.
If branching rate is turned on, however,
a spontaneous annihilation of a single particle by the chain of reactions
$A \rightarrow 2A \rightarrow 0$ can occur, which results in
an exponential density decay.  That is, the BAW$_1$ with $p=1$ cannot capture the main feature 
(exponential density decay in this example) of its absorbing phase.

\begin{figure}
	\includegraphics[width=\linewidth]{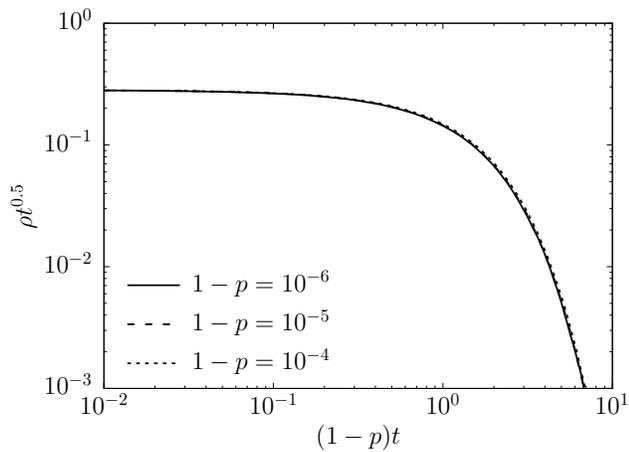}
	\caption{\label{Fig:bw1} Scaling-collapse plot of $\rho t^{0.5}$ vs $(1-p)t$
	of the BAW$_1$ for $1-p=10^{-4}$, $10^{-5}$, and $10^{-6}$ on a double-logarithmic
	scale. As anticipated by Eq.~\eqref{Eq:bwc}, curves for different $p$'s 
	are hardly discernible.
	}
\end{figure}
Actually, the behavior of the BAW$_1$ around $p=1$ can be described by a scaling function
\begin{align}
	\rho(t) = t^{-0.5} F[(1-p)t],
	\label{Eq:bwc}
\end{align}
where $F(x)$ is expected to decrease exponentially for large $x$.
The reason why $(1-p)t$ should be a single scaling parameter is clear.
The spontaneous annihilation can be crucial only when substantial amount of branching events have occurred,
which is expected if time elapses more than $1/(1-p)$.
In Fig.~\ref{Fig:bw1}, we show scaling collapse of the BAW$_1$ for 
$p$ close to 1, which confirms the scaling ansatz~\eqref{Eq:bwc}. 
Here, the system size is $2^{25}$ and average over 8 independent runs
for each parameter is taken. As the example of the BAW$_1$ reveals, it is possible that
$p=1$ of the modified model is in a sense a singular point and
that the modified model in the absorbing phase does not exhibit $t^{-0.5}$ 
behavior for small $\sigma$.

\begin{figure}
	\includegraphics[width=\linewidth]{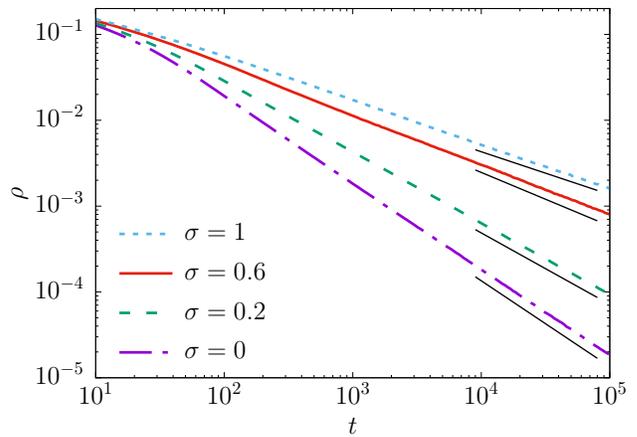}
	\caption{\label{Fig:brb} Double-logarithmic plots of $\rho$ vs $t$
	of the model with dynamic rules \eqref{Eq:modi} for $\sigma = 0$,
	0.2, 0.6, and 1 (bottom to top). Here, we set $p=0.8$ for all cases.
	For guides to the eyes, we also depict a line segment with slope
	$-1/(1+\sigma)$ right below each curve.
	}
\end{figure}

To obtain the answer, we now resort to Monte Carlo simulations.
Using systems of size $L=2^{24}$, we performed simulations 
for $\epsilon=0.5$ and $p=0.8$. To reduce statistical error, we performed 40 independent runs for each parameter set.
Figure~\ref{Fig:brb} shows the behavior of the density for $\sigma=0$, 0.2, 0.6, and 1 
on a double logarithmic scale. 
Just like the BAWL with $p=1$, the density decays as $t^{-\alpha}$
with $\alpha$ in  Eq.~\eqref{Eq:al}.
Hence, we expect that the critical behavior is the same regardless of
whether hopping or branching is biased. We have checked this anticipation
by simulations and our preliminary simulations for $\sigma=0$ indeed
show that the critical behavior of the modified model is the same as the BAWL (details not shown here). 
This also indirectly confirms that the BAWL with $p=1$ 
correctly represents the behavior in the absorbing phase.
To conclude this section, we have shown that 
the presence of the local bias due to 
long-range attraction is enough to exhibit non-DI critical phenomena,
irrespective of which dynamic process the local bias is applied.

\section{\label{Sec:sum}Summary}
To summarize, we studied the branching annihilating random walks
with long-range attraction (BAWL). The long-range attraction
has a power-law feature with exponent $\sigma$; see Eq.~\eqref{Eq:q}. 
We investigated the critical decay exponent $\delta$ that describes how the density
behaves with time at the critical point.
We first numerically found that $\delta$ varies continuously with $\sigma$ for $\sigma<1$ and
is the same as the critical decay exponent of the directed Ising universality class
for $\sigma \ge 1$. 
By the analysis of a random walk with an attracting center at the origin
together with Monte Carlo simulations for the BAWL with $p=1$,
we argued that $\sigma_c$ should be 1.

We also studied the modified model in which offspring prefer being placed toward the nearest particle
but hopping is now unbiased. 
We found that the absorbing phase of the modified model shows the
same asymptotic behavior of  the BAWL for the same value of $\sigma$. Therefore, we concluded that
it is immaterial which dynamic process, hopping or branching, is biased by
the long-range attraction.

\begin{acknowledgments}
	This work was supported by the Basic Science Research Program through the
National Research Foundation of Korea~(NRF) funded by the Ministry of
Science and ICT~(Grant No. 2017R1D1A1B03034878).
The author furthermore thanks the Regional Computing Center of the
University of Cologne (RRZK) for providing computing time on the DFG-funded High
Performance Computing (HPC) system CHEOPS.
\end{acknowledgments}
\bibliography{Park}
%\bibliography{../../prebib.bib}
\end{document}

\begin{CJK*}{UTF8}{mj}
\title{Supplemental Material for ``Branching Annihilating Random Walks with Long-Range Attraction in One Dimension''}
\author{Su-Chan Park (박수찬)}
\affiliation{The Catholic University of Korea, Bucheon 14662, Republic of Korea}
\begin{abstract}
In this Supplemental Material, we present details of numerical analyses for 
the corrections-to-scaling function $Q$ and the effective exponent $\delta_\text{e}$ that 
are defined in the main text.
Since the coefficients of $Q$ for all cases are found to be negative, we depict 
$-Q$ as a function of $t$ on a double logarithmic scale. We estimate $\chi$ by observing that
the asymptotic behavior of $Q$ does not depend on $b$.
We also explain how we find the critical point from the effective exponent.
\end{abstract}
\maketitle
\end{CJK*}

\begin{center}
\includegraphics[width=0.95\linewidth]{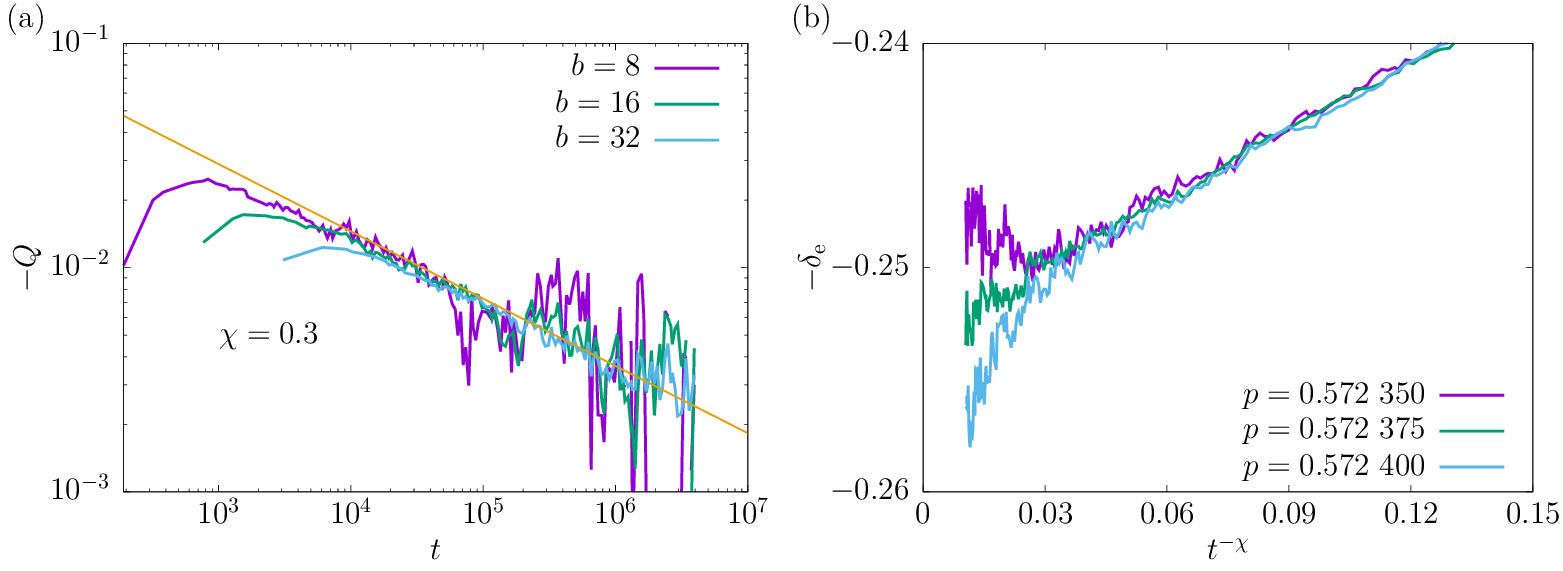}
\end{center}

\noindent
Fig. S1. $\sigma=0.1$. (a) Double-logarithmic plots of $-Q$ vs $t$ at the critical point with $b=8$, 16, and 32. The estimated $\chi$ is explicitly written. We also depicts $t^{-\chi}$ for guides to the eyes. (b) Plots of $-\delta_\text{e}$ vs $t^{-\chi}$ around the critical point with $b=16$. 
Since the middle curve shows the best linear behavior, we conclude
$p_c=0.572~375(25)$.

\begin{center}
\includegraphics[width=0.95\linewidth]{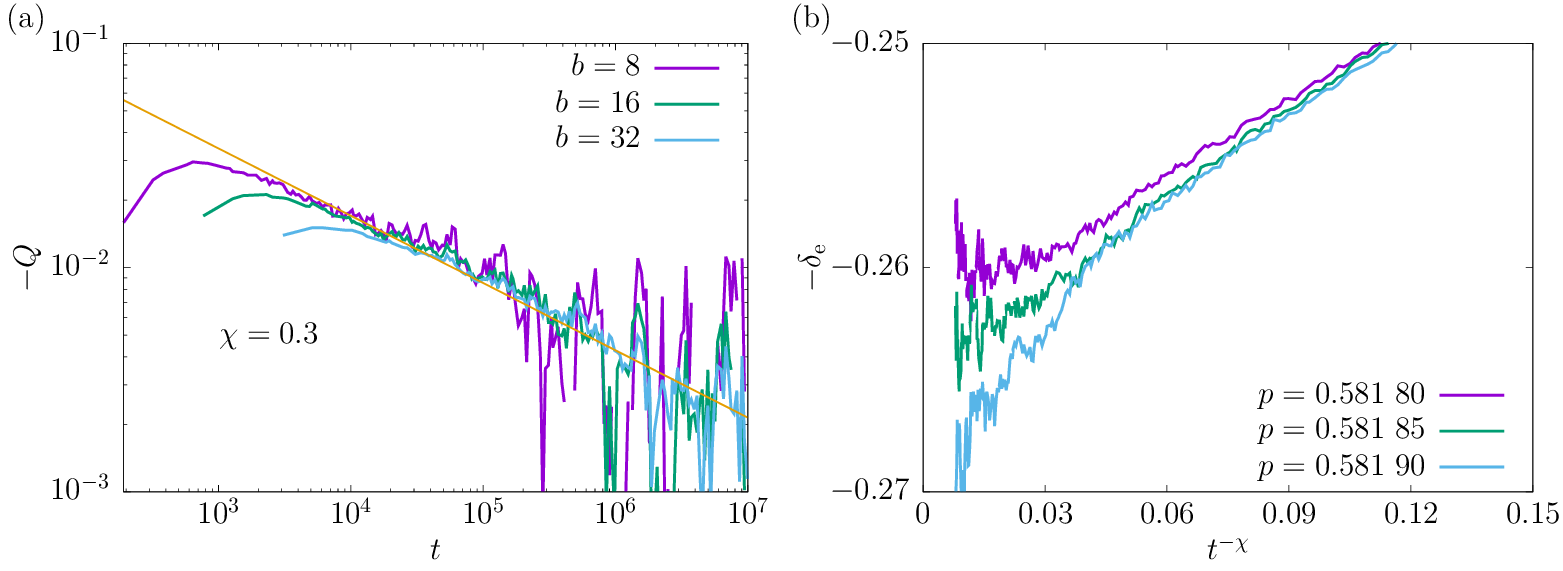}
\end{center}

\noindent
Fig. S2. $\sigma=0.2$. (a) Double-logarithmic plots of $-Q$ vs $t$ at the critical point with $b=8$, 16, and 32. The estimated $\chi$ is explicitly written. We also depicts $t^{-\chi}$ for guides to the eyes. (b) Plots of $-\delta_\text{e}$ vs $t^{-\chi}$ around the critical point with $b=16$. 
Since the middle curve shows the best linear behavior, we conclude
$p_c=0.581~85(5)$.

\begin{center}
\includegraphics[width=0.95\linewidth]{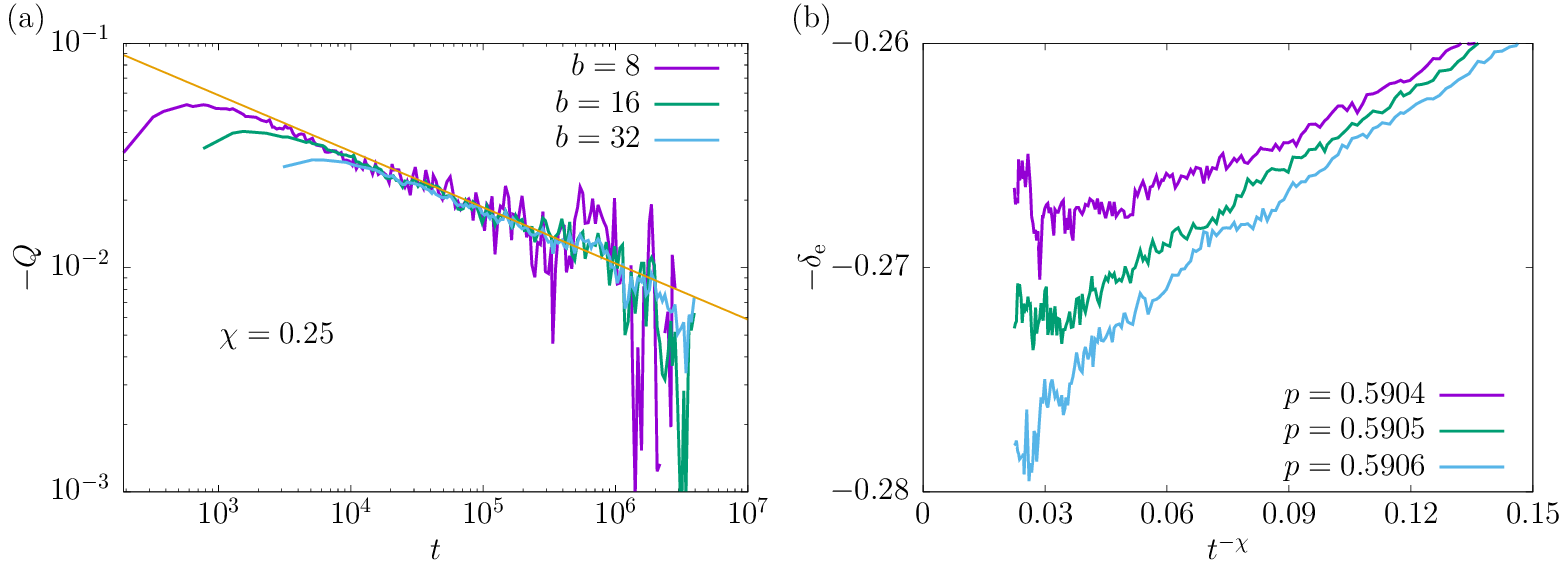}
\end{center}

\noindent
Fig. S3. $\sigma=0.3$. (a) Double-logarithmic plots of $-Q$ vs $t$ at the critical point with $b=8$, 16, and 32. The estimated $\chi$ is explicitly written. We also depicts $t^{-\chi}$ for guides to the eyes. (b) Plots of $-\delta_\text{e}$ vs $t^{-\chi}$ around the critical point with $b=16$. 
Since the middle curve shows the best linear behavior, we conclude
$p_c=0.5905(1)$.

\begin{center}
\includegraphics[width=0.95\linewidth]{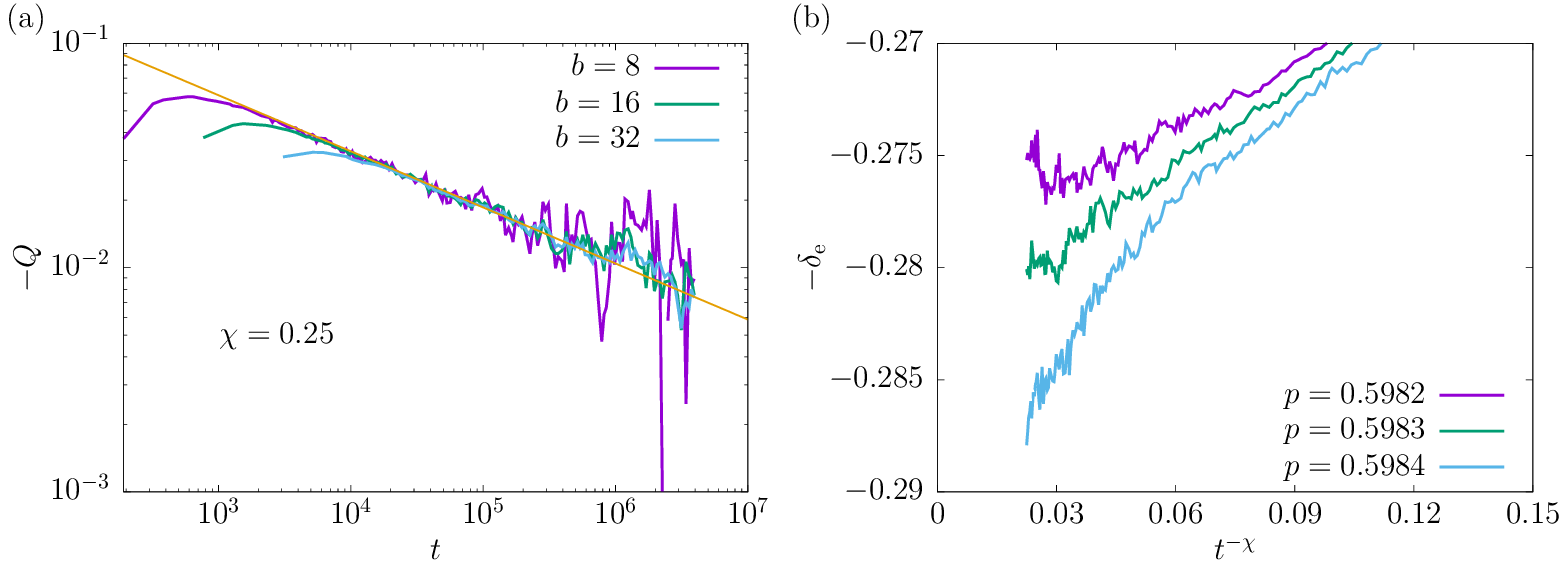}
\end{center}

\noindent
Fig. S4. $\sigma=0.4$. (a) Double-logarithmic plots of $-Q$ vs $t$ at the critical point with $b=8$, 16, and 32. The estimated $\chi$ is explicitly written. We also depicts $t^{-\chi}$ for guides to the eyes. (b) Plots of $-\delta_\text{e}$ vs $t^{-\chi}$ around the critical point with $b=16$. 
Since the middle curve shows the best linear behavior, we conclude
$p_c=0.5983(1)$.

\begin{center}
\includegraphics[width=0.95\linewidth]{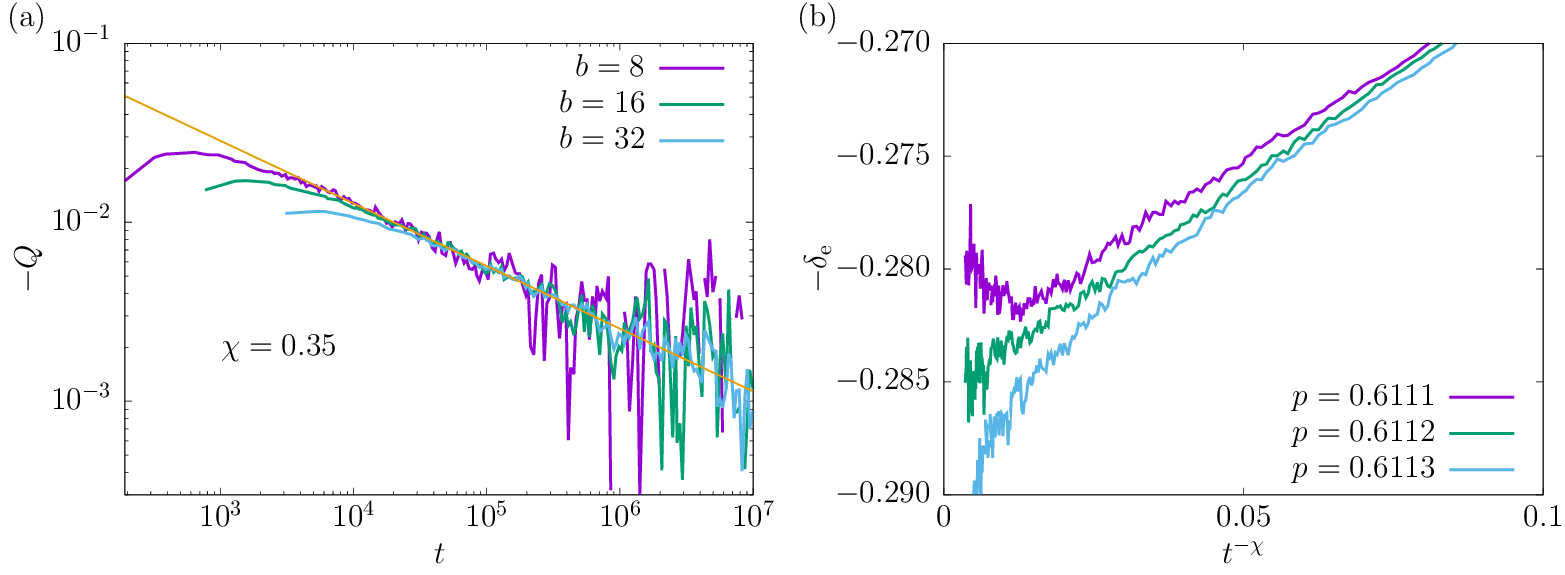}
\end{center}

\noindent
Fig. S5. $\sigma=0.6$. (a) Double-logarithmic plots of $-Q$ vs $t$ at the critical point with $b=8$, 16, and 32. The estimated $\chi$ is explicitly written. We also depicts $t^{-\chi}$ for guides to the eyes. (b) Plots of $-\delta_\text{e}$ vs $t^{-\chi}$ around the critical point with $b=16$. 
Since the middle curve shows the best linear behavior, we conclude
$p_c=0.6112(1)$.

\begin{center}
\includegraphics[width=0.95\linewidth]{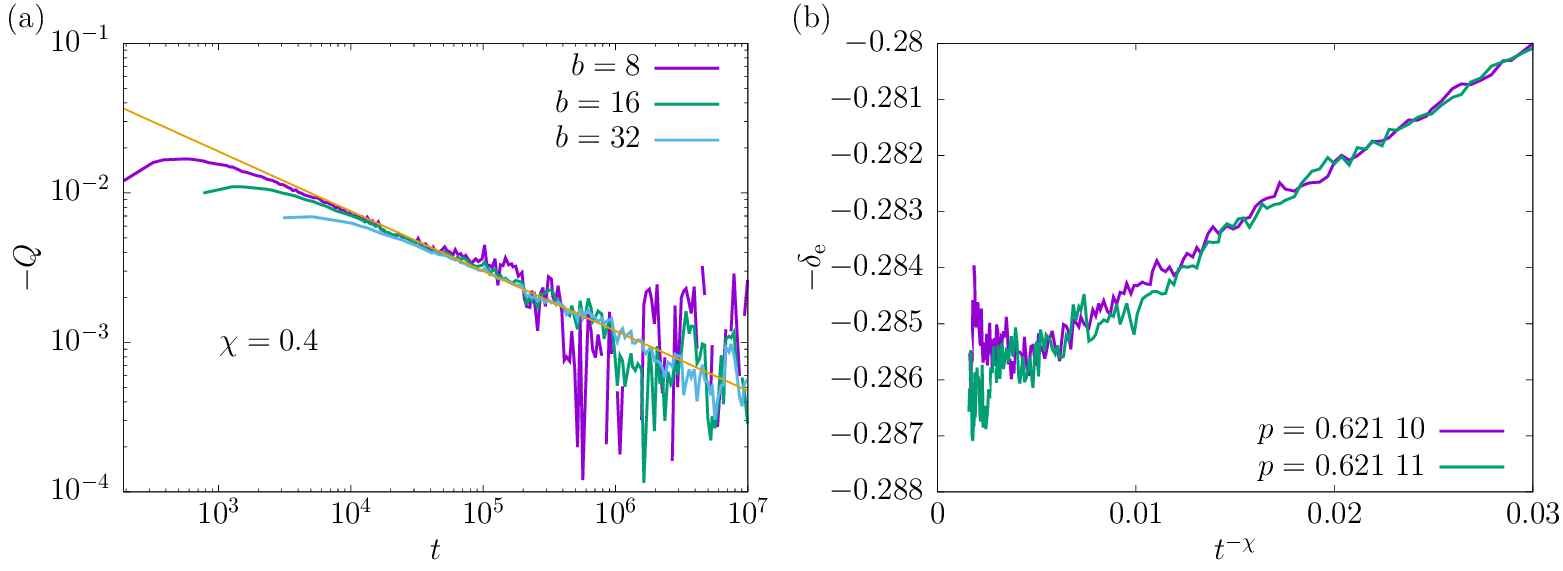}
\end{center}

\noindent
Fig. S6. $\sigma=0.8$. (a) Double-logarithmic plots of $-Q$ vs $t$ at the critical point with $b=8$, 16, and 32. The estimated $\chi$ is explicitly written. We also depicts $t^{-\chi}$ for guides to the eyes. (b) Plots of $-\delta_\text{e}$ vs $t^{-\chi}$ around the critical point with $b=16$. 
Since the bottom curve shows the best linear behavior, we conclude
$p_c=0.621~11(1)$.

\begin{center}
\includegraphics[width=0.95\linewidth]{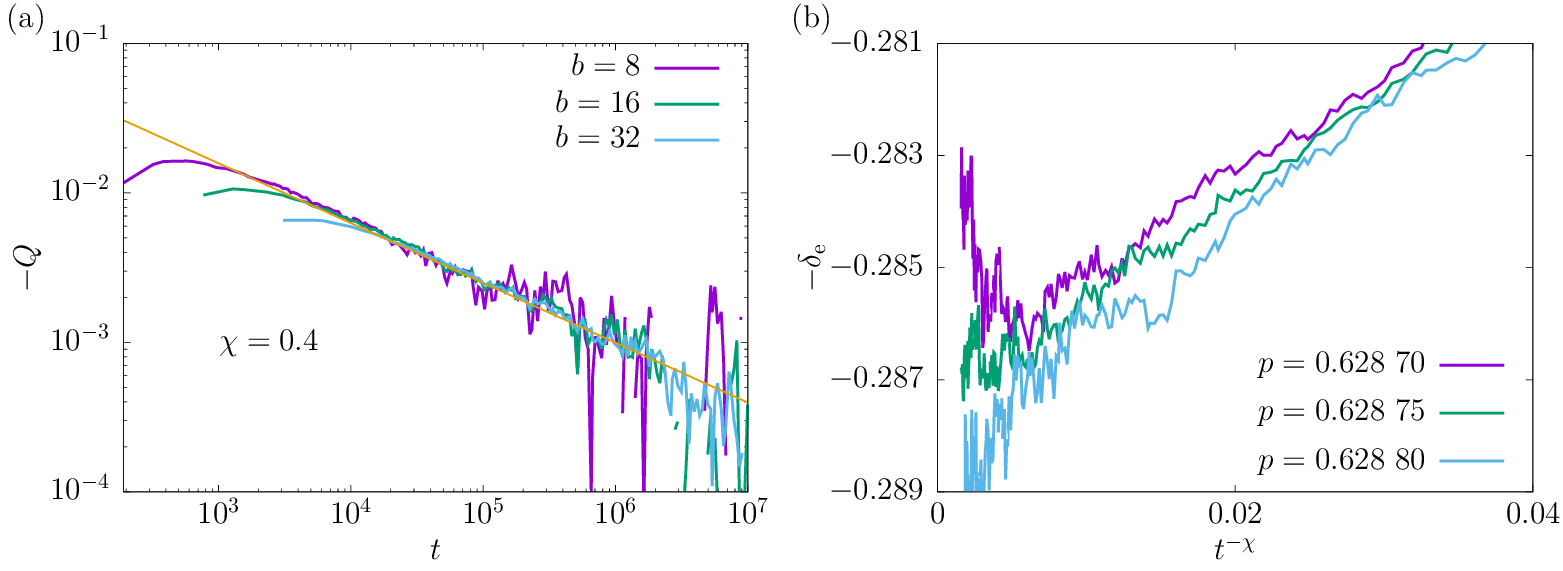}
\end{center}

\noindent
Fig. S7. $\sigma=1$. (a) Double-logarithmic plots of $-Q$ vs $t$ at the critical point with $b=8$, 16, and 32. The estimated $\chi$ is explicitly written. We also depicts $t^{-\chi}$ for guides to the eyes. (b) Plots of $-\delta_\text{e}$ vs $t^{-\chi}$ around the critical point with $b=16$. 
Since the middle curve shows the best linear behavior, we conclude
$p_c=0.628~75(5)$.